\documentclass[12pt]{iopart}

\usepackage{amsmath,amsthm,amsfonts,amssymb,bm,epsfig}

\begin{document}

\title[Optimal control of quantum gates and suppression of
  decoherence]{Optimal control of quantum gates and suppression of
  decoherence in a system of interacting two-level particles}

\author{Matthew~Grace$^1$, Constantin~Brif$^1$, Herschel~Rabitz$^1$,
  Ian~A.~Walmsley$^2$, Robert~L.~Kosut$^3$, and Daniel~A.~Lidar$^4$}

\address{$^1$ Department of Chemistry, Princeton University, Princeton,
  New Jersey 08544}

\address{$^2$ Department of Physics, University of Oxford, Oxford OX1
  3PU, UK}

\address{$^3$ SC Solutions, Inc., 1261 Oakmead Parkway, Sunnyvale, CA
  94085}

\address{$^4$ Departments of Chemistry, Electrical Engineering, and
  Physics, University of Southern California, Los Angeles, CA 90089}

\eads{\mailto{mgrace@princeton.edu}, \mailto{cbrif@princeton.edu},
  \mailto{hrabitz@princeton.edu}, \mailto{walmsley@physics.ox.ac.uk},
  \mailto{kosut@scsolutions.com}, and \mailto{lidar@usc.edu}}

\date{\today}

\begin{abstract}
Methods of optimal control are applied to a model system of interacting
two-level particles (e.g., spin-half atomic nuclei or electrons or
two-level atoms) to produce high-fidelity quantum gates while
simultaneously negating the detrimental effect of decoherence. One set
of particles functions as the quantum information processor, whose
evolution is controlled by a time-dependent external field. The other
particles are not directly controlled and serve as an effective
environment, coupling to which is the source of decoherence. The control
objective is to generate target one- and two-qubit unitary gates in the
presence of strong environmentally-induced decoherence and under
physically motivated restrictions on the control field. The quantum-gate
fidelity, expressed in terms of a novel state-independent distance
measure, is maximized with respect to the control field using combined
genetic and gradient algorithms. The resulting high-fidelity gates
demonstrate the feasibility of precisely guiding the quantum evolution
via optimal control, even when the system complexity is exacerbated
by environmental coupling. It is found that the gate duration has an
important effect on the control mechanism and resulting fidelity. An
analysis of the sensitivity of the gate performance to random variations
in the system parameters reveals a significant degree of robustness
attained by the optimal control solutions.
\end{abstract}


\submitto{\jpb}

\maketitle


\section{Introduction}

The transfer of information between elements of a quantum computational
system requires the use of entangling quantum interactions
\cite{NC00}. Undesired interactions between the system and its
surroundings can destroy quantum coherences and thus are a critical
obstacle to successful quantum computation (QC). The feasibility of
creating high-fidelity quantum gates in the presence of
environmentally-induced decoherence is one of the most important
problems to overcome for practical QC. In particular, in spin-based
solid-state realizations of QC \cite{Loss, Kane, Vrijen, Petta} one
encounters a difficult task of effectively separating a multiparticle
quantum system into interacting and non-interacting components.

Quantum error correction (QEC) enables fault-tolerant QC \cite{FTQC},
but only when the errors in quantum gate operations are sufficiently
small \cite{Threshold}. Therefore, it is very important to decrease the
errors caused by decoherence. This problem has inspired significant
interest in various methods of decoherence management, including the use
of decoherence-free subspaces and noiseless subsystems \cite{PSE96,
DuGu97, ZaRa97, LiWh, KLV00}, quantum dynamical decoupling \cite{ViLl,
Zan99, ViTo, UA02, LiWuBy, KoKu04, Fac05, FHN06}, schemes based on
stochastic control \cite{Stoch}, optimal control techniques
\cite{Brif01, ZhRa03, HoSt04, STK04, GrKh05, Potz}, and multilevel
encoding of logical states \cite{MLE}.

The method of optimal control \cite{RVM00, WaRa03} enables managing the
dynamics of complex quantum systems in a very precise and specific
manner and therefore is especially useful in QC. In addition to
applications to the problem of dynamical suppression of decoherence
\cite{Brif01, ZhRa03, HoSt04, STK04, GrKh05, Potz}, optimal control
theory (OCT) \cite{PeDaRa, WRD93} was also successfully used to design
unitary quantum gates in closed systems \cite{SKH99, PaKo, SkTa,
Khaneja05, Sp07}. The optimal control of quantum gates in the presence
of decoherence still remains to be fully explored. In \cite{MLE} we
previously considered the optimal control of quantum gates for qubits
encoded in multilevel subspaces; this method makes quantum gates immune
to mixing and decoherence that occur within the encoding
subspaces. Recent works \cite{GK05, Schulte06} developed specific
techniques, involving optimizations over sets of controls operating in
pre-designed ``weak-decoherence'' subspaces. In the present paper we
propose a different approach in which the full power of OCT is used to
generate the target gate with the highest possible fidelity while
simultaneously suppressing strong decoherence induced by coupling to a
multiparticle environment. This method does not rely on any special
pre-design of the system parameters to avoid or weaken decoherence
(e.g., using multiple levels as in \cite{MLE}, tunable inter-qubit
couplings as in \cite{GK05}, or auxiliary qubits as in
\cite{Schulte06}); the only control used in the present approach is a
time-dependent external field.

A similar OCT-based approach was recently used \cite{Hoh06} to design
quantum gates for solid-state qubits in the presence of decoherence.
However, the objective in \cite{Hoh06} was to optimize a
purity-dependent quality factor (or, in \cite{GK05}, the purity itself),
instead of the actual gate fidelity. In the present work we demonstrate
that although improving the purity of the quantum information processor
(QIP) is necessary for performing a high-fidelity quantum gate, it is
not sufficient. Even if the QIP is completely decoupled from its
environment at a given time, this does not ensure that the desired gate
operation will be performed at the decoupling time. Therefore, we
optimize a gate fidelity \cite{KGBR06} which directly measures the
distance between the target quantum gate of the QIP and the actual
transformation of the composite system. Optimization techniques were
also applied recently to QEC \cite{ReWe05, KoLi06}. In contrast to QEC,
our approach does not require ancilla qubits and is not limited to the
weak decoherence regime. The optimal control of quantum gates can
potentially be used in conjunction with QEC to achieve fault tolerance
with an improved threshold.

In this work, we consider a model system composed of interacting
two-level particles, for example, spin-half atomic nuclei or electrons
or two-level atoms. A small set of particles serve as qubits in the
QIP; the rest of the particles serve as an effective environment. The
qubits are directly controlled by a time-dependent external field, while
the environmental particles do not directly couple to the field. The
control objective is to generate target quantum gates in the QIP with
the highest possible fidelity. The optimal control field must perform
the desired gate operation while simultaneously suppressing the
qubit-environment interaction and restoring lost coherence to the
QIP. This model is sufficiently simple to allow for a full numerical
treatment of the entire composite system, and the results are relevant
to important physical applications, in particular, to spin-based
solid-state realizations of quantum gates \cite{Loss, Kane, Vrijen,
Petta}. For example, our model bears a similarity to systems in which an
electron spin (or a pair of electron spins) is coupled to a nuclear spin
bath \cite{Petta, Chil, Coish}. Coherent manipulation of electron spins
via rapid electrical control of the exchange interaction has been
successfully demonstrated in such systems \cite{Petta}. The analysis
reported in the present work indicates that the employment of the
optimal control methods may increase the effectiveness of coherent
management of coupled spin dynamics.

The paper is organized as follows. Section~\ref{sec:model} presents the
model (including an explicit matrix form for the simplest case of one
qubit coupled to a one-particle environment) and schemes of
multiparticle couplings. In section~\ref{sec:dist}, we consider a
distance measure that quantifies the fidelity of quantum gates. This
fidelity is independent of the initial state and is evaluated directly
from the evolution operator of the composite system.
Section~\ref{sec:deco} investigates the dynamics of decoherence in the
uncontrolled system for various values of system parameters. In order to
fully explore the utility of OCT, we select a set of parameters that
enhances the loss of coherence in the uncontrolled system. In
section~\ref{sec:optimize}, we describe in detail the genetic and
gradient optimization algorithms. The results obtained with the optimal
controls are presented and discussed in
section~\ref{sec:results}. Section~\ref{sec:robust} investigates the
robustness of optimal solutions to uncertainties in the system
parameters. Finally, section~\ref{sec:conc} concludes with a summary of
the results and discusses future directions.


\section{The model system}
\label{sec:model} 

We use a model of $N$ interacting two-level particles (e.g., spin-half
particles or two-level atoms), which are divided into the QIP, composed
of $m$ qubits, and an $n$-particle environment ($N = m + n$). The qubits
are directly coupled to a time-dependent external control field, while
the environment is not directly controlled and is managed only through
its interaction with the qubits. The evolution of the composite system
of qubits and environment is treated in an exact quantum-mechanical
manner, without either approximating the dynamics by a master equation
or using a perturbative analysis based on the weak coupling
assumption. The Hamiltonian for the composite controlled system, $H =
H_0 + H_C + H_{\mathrm{int}}$, has the form $(\hbar = 1)$
\begin{equation}
H = \sum_{i = 1}^N \omega_i S_{i z} 
- \sum_{i = 1}^{m} \mu_i C(t) S_{i x}
- \sum_{i = 1}^{N - 1} \sum_{j > i}^N \gamma_{i j} \mathbf{S}_i
\cdot \mathbf{S}_j.
\label{Ham}
\end{equation}
Here, $\mathbf{S}_i = \left( S_{i x}, S_{i y}, S_{i z} \right)$ is the
spin operator for the $i$th particle ($\mathbf{S}_i = \frac{1}{2}
\mbox{\boldmath$\sigma$}_i$, in terms of the Pauli matrices), $H_0$ is
the sum over the free Hamiltonians $\omega_i S_{i z}$ for all $N$
particles ($\omega_i$ is the transition angular frequency for the $i$th
particle), $H_C$ specifies the coupling between the $m$ qubits and the
time-dependent control field $C(t)$ ($\mu_i$ are the dipole moments),
and $H_{\mathrm{int}}$ represents the Heisenberg exchange interaction
between the particles ($\gamma_{i j}$ is the coupling constant for the
$i$th and $j$th particles). This model is particularly relevant to
spin-based solid-state realizations of quantum gates
\cite{Loss, Kane, Vrijen, Petta}.

Now consider the simplest case of one qubit and a one-particle
environment $(m = n = 1)$ in more detail. The Hamiltonian in this case
is:
\begin{equation}
H = \omega_1 S_{1 z} + \omega_2 S_{2 z} - \mu C(t) S_{1 x} - \gamma
\mathbf{S}_1 \cdot \mathbf{S}_2,
\label{Ham_11}
\end{equation}
where $\gamma = \gamma_{12}$. We use the orthonormal basis: 
\begin{equation}
|1\rangle = |+\rangle_1 \otimes |+\rangle_2, \ \ 
|2\rangle = |+\rangle_1 \otimes |-\rangle_2, \ \ 
|3\rangle = |-\rangle_1 \otimes |+\rangle_2, \ \ 
|4\rangle = |-\rangle_1 \otimes |-\rangle_2,
\label{basis}
\end{equation}
where $S_{i z} |\pm\rangle_i = \pm \frac{1}{2} |\pm\rangle_i$. The
Hamiltonian (\ref{Ham_11}) in the basis (\ref{basis}) has the
following matrix form:
\begin{equation}
H = \frac{1}{2} \left(
\begin{array}{cccc}
\omega_1 + \omega_2 - \frac{1}{2} \gamma & 0 & -\mu C(t) & 0 \\
0 & \omega_1 - \omega_2 + \frac{1}{2} \gamma & -\gamma & -\mu C(t) \\
-\mu C(t) & -\gamma & \omega_2 - \omega_1 + \frac{1}{2} \gamma & 0 \\
0 & -\mu C(t) & 0 & - \omega_1 - \omega_2 - \frac{1}{2} \gamma
\end{array} \right).
\end{equation}

In addition to the simplest case of a two-particle system described
above, we also consider situations where one qubit is coupled to a
multiparticle environment ($m = 1$ and $n = 2, 4, 6$). For $m = 1$, the
coupling constants are given by
\begin{equation}
\gamma_{i j} = \left\{ \begin{array}{ll} \gamma, 
& \mathrm{for} \ i = 1 \ \mathrm{and} \ j = 2, \ldots, N, \\ 
0, & \mathrm{for} \ 2 \leq i \leq N,
\end{array}\right.
\label{gamma}
\end{equation}
which means that the qubit interacts with each environmental particle
with the same coupling constant $\gamma$, and the environmental
particles are not directly coupled to each other. For $n = 2$, the
system can be modeled by a linear chain with the qubit $q_1$ at the
center, equally coupled to both environmental particles $e_2$ and $e_3$:
\begin{equation}
e_2 \longleftrightarrow q_1 \longleftrightarrow e_3
\end{equation}
For $n = 4$, the system can be modeled by a two-dimensional lattice with
the qubit $q_1$ at the center, equally coupled to four environmental
particles $\{ e_2 , \ldots , e_5 \}$:
\begin{equation}
\begin{array}{c}
e_4 \\ \updownarrow \\ 
e_2 \longleftrightarrow q_1 \longleftrightarrow e_3 \\
\updownarrow \\ e_5
\end{array}
\label{square_c}
\end{equation}
Similarly, for $n = 6$, the system can be modeled by a three-dimensional
lattice with the qubit at the center, coupled to six environmental
particles. In these lattices, it is assumed that the Heisenberg
interactions decay exponentially with distance \cite{Loss}, and
therefore environmental particles on the vertices of the square $(n =
4)$ and cube $(n = 6)$ are neglected.

A different model with nearest-neighbor couplings is also considered in
the case of $n = 4$. The system is modeled by a linear chain of
particles, with the qubit at the center and each particle coupled only
to its nearest neighbors with the same coupling constant $\gamma$:
\begin{equation}
e_4 \longleftrightarrow e_2 \longleftrightarrow q_1 
\longleftrightarrow e_3 \longleftrightarrow e_5
\label{nnc} 
\end{equation}

The case where two qubits are coupled to a one-particle environment ($m
= 2$ and $n = 1$) is used to develop an entangling quantum gate
(specifically, the controlled-NOT gate) in the presence of a simple
environment. This system can be modeled by the following two-dimensional
triangular lattice:
\begin{equation}
\begin{array}{ccc}
& e_3 & \\ 
\stackrel{\gamma_{13}}{} \swarrow \! \! \! \! \! \! \nearrow & &
\nwarrow \! \! \! \! \! \! \searrow \stackrel{\gamma_{23}}{} \\ 
q_1 \ \ & \stackrel{\gamma_{12}}{\longleftrightarrow} & \ \ q_2
\end{array}
\end{equation}
where the two qubits are denoted as $q_1$ and $q_2$, and the
environmental particle as $e_3$. Such a model is relevant, for example,
for a dilute nuclear spin bath \cite{Kane}. Values for this set of
coupling constants are given in section~\ref{sec:CNOT}.


\section{The distance measure}
\label{sec:dist}

Our objective is to generate an evolution of the QIP which at some time
$t_{\mathrm{f}}$ will be as close as possible to the target quantum
gate. The problem of evaluating the actual gate fidelity is complicated
by the fact that the evolution of the QIP is non-unitary due to the
interaction with the environment. Nevertheless, it is possible to define
a useful measure of the distance between the target quantum gate of the
QIP and the actual evolution operator of the composite system
\cite{KGBR06}.

Let $U(t) \in \mathrm{U}(2^N)$ be the unitary time-evolution operator of
the composite system and $G \in \mathrm{U}(2^m)$ be the unitary target
transformation for the quantum gate of the QIP (where $\mathrm{U}(d)$
denotes the group of all $d \times d$ unitary matrices). The evolution
of the composite system is governed by the Schr\"{o}dinger equation,
\begin{equation}
\dot{U}(t) = - \rmi H(t) U(t),
\label{Schro}
\end{equation}
with the initial condition $U(0) = I_{2^N}$ (where $I_d$ denotes the $d
\times d$ identity matrix). The gate fidelity depends on the distance
between the actual evolution $U \equiv U(t_{\mathrm{f}})$ at the final
time $t_{\mathrm{f}}$ and the target transformation $G$. In order to
perform a perfect gate, it suffices for the time-evolution operator at
$t = t_{\mathrm{f}}$ to be in a tensor-product form $U_{\mathrm{opt}} =
G \otimes \Phi$, where $\Phi \in \mathrm{U}(2^n)$ is an arbitrary
unitary transformation acting on the environment.\footnote{We do not
consider in the present work a more general situation where the
composite system itself is open and $\Phi$ may not be unitary.}
Therefore, the following objective functional is proposed \cite{KGBR06}
as the measure of the distance between $U$ and $G$:
\begin{equation}
J = \lambda_N \underset{\Phi}{\min} \left\{ \| U - G \otimes \Phi \| \
  \left| \ \Phi \in \mathrm{U}(2^n) \right. \right\},
\label{distance}
\end{equation}
where $\| \cdot \|$ is a matrix norm on the space $M_d \left( \mathbb{C}
\right)$ of $d \times d$ complex matrices (in the present case $d =
2^N$), $\lambda_N$ is a normalization factor, and $J$ is minimized over
the set of all unitary $\Phi$. It is useful to expand $G$, $\Phi$, and
$U$ in orthonormal bases. Let $\{ |i\rangle \}$, $\{ |\nu\rangle \}$ and
$\{ |i\rangle \otimes |\nu\rangle \}$ be orthonormal bases that span the
Hilbert spaces of the QIP, environment, and composite system,
respectively. The corresponding expansions read
\begin{subequations}
\label{mas}
\begin{eqnarray}
& & G = \sum_{i, i' = 1}^{2^m} G_{i i'} 
| i \rangle\langle i' |, \ \ \ 
\Phi = \sum_{\nu, \nu' = 1}^{2^n} \Phi_{\nu \nu'} 
| \nu \rangle\langle \nu' |,
\label{target} \\
& & U = \sum_{i , i' = 1}^{2^m} \sum_{\nu , \nu' = 1}^{2^n} 
U_{\substack{i i' \\ \nu \nu' }} 
| i \rangle\langle i' | \otimes | \nu \rangle\langle \nu' |.
\label{Utensor}
\end{eqnarray}
\end{subequations}

Using in (\ref{distance}) the Frobenius norm, defined as
\begin{equation}
\| X \|_{\mathrm{Fr}} = \left[ \Tr \left( X^{\dag} X \right)
\right]^{1/2} \ \ \ \forall X \in M_{d} \left( \mathbb{C} \right),
\label{Frobnorm}
\end{equation}
and $\lambda_N = 2^{-(N + 1)/2}$, the distance measure becomes
\cite{KGBR06}
\begin{equation}
J = \left[ 1 - 2^{-N} \Tr \left( \sqrt{Q^{\dag}Q} \right) \right]^{1/2},
\label{frobdist1}
\end{equation}
where $Q \in M_{2^n}(\mathbb{C})$ is given by
\begin{equation}
Q = \sum_{\nu, \nu' = 1}^{2^n} \left( \sum_{i, i' = 1}^{2^m} G_{i
i'}^{\ast} U_{\substack{i i' \\ \nu \nu'}}, \right) | \nu
\rangle\langle \nu' |.
\label{Q-def}
\end{equation}
Since $0 \leq J \leq 1$, it is convenient to define the gate fidelity as
$F = 1 - J$. An important property of this distance measure is its
independence of the initial state. In contrast to some other distance
measures,\footnote{Relationships between various distance measures,
including some presented in \cite{GLN05} and generalizations of
(\ref{frobdist1}), are discussed in more detail in \cite{KGBR06}.} $J$
is evaluated directly from the evolution operator $U$, with no need to
specify the initial state of the system. This property of $J$ reflects
our objective of generating a specified target transformation for
whatever initial state, pure or mixed, direct-product or entangled.

Note that in the ideal case when there is no coupling to the
environment, i.e., the QIP is a closed system with unitary dynamics,
the distance measure (\ref{frobdist1}) becomes
\begin{equation}
J = \left[ 1 - 2^{-m} \left| \Tr \left( G^{\dag} U_{\mathrm{q}}
  \right) \right| \right]^{1/2} ,
\end{equation}
where $U_{\mathrm{q}} \equiv U_{\mathrm{q}}(t_{\mathrm{f}})$ is the
unitary evolution operator of the QIP at the final time. Another
distance measure used in the literature \cite{Sp07} for closed systems
is $J_{\mathrm{cs}} = 1 - 2^{-m} \left| \Tr ( G^{\dag} U_{\mathrm{q}} )
\right|$, i.e., $J_{\mathrm{cs}} = J^2$. For example, in
section~\ref{sec:results} we report optimization results which, in the
case of closed QIP systems, are $J \sim 10^{-6}$ and $J \sim 10^{-4}$
for one- and two-qubit gates, respectively, corresponding to the values
$J_{\mathrm{cs}} \sim 10^{-12}$ and $J_{\mathrm{cs}} \sim 10^{-8}$,
respectively.


\section{Decoherence dynamics of the uncontrolled system}
\label{sec:deco}

The loss of coherence in the QIP, caused by the interaction with the
environment, is detrimental to the quantum gate performance. In order to
better understand the mechanism of optimal control, we first study
the decoherence process in the uncontrolled system. The state of the QIP
at time $t$ is described by the reduced density matrix:
\begin{equation}
\rho_{\mathrm{q}}(t) = \Tr_{\mathrm{env}} \left[ \rho(t) \right],
\label{red-dens}
\end{equation}
where $\rho(t)$ is the density matrix of the composite system and
$\Tr_{\mathrm{env}}$ denotes the trace over the environment. A useful
measure of decoherence is the von Neumann entropy \cite{PK91}:
\begin{equation}
S_{\mathrm{vN}}(t) = - \Tr \left\{ \rho_{\mathrm{q}}(t) \ln \left[
  \rho_{\mathrm{q}}(t) \right] \right\}.
\label{vN}
\end{equation}
For a pure state, $S_{\mathrm{vN}} = 0$, while for a maximally mixed
state of a $k$-level system, $S_{\mathrm{vN}} = \ln(k)$. We explore the
decoherence dynamics of the QIP by studying the time evolution of the
entropy $S_{\mathrm{vN}}(t)$ for the uncontrolled system (in this
section) and under the influence of optimal time-dependent control
fields (in subsequent sections). The initial state used for the entropy
calculations is
\begin{equation}
|\Psi_0 \rangle = \bigotimes_{i = 1}^{m} |-\rangle_i \otimes
\bigotimes_{j = m + 1}^{N} |+\rangle_j
\label{initial}
\end{equation}
(i.e., initially all qubits are in the state $|-\rangle$ and all
environmental particles are in the state $|+\rangle$). Recall that the
distance measure $J$ of (\ref{frobdist1}) is independent of the initial
state and consequently so are the optimal control fields found for the
target gates and the corresponding fidelities. We choose some initial
state only for the entropy calculations, which are done to illustrate
the decoherence dynamics \emph{after} the time-evolution operator is
determined (for either a controlled or uncontrolled system). Therefore,
the specific choice of the initial state (\ref{initial}) places no
limitations whatsoever on the generality of the optimal control results.

\begin{figure}
\epsfxsize=0.7\textwidth \centerline{\epsffile{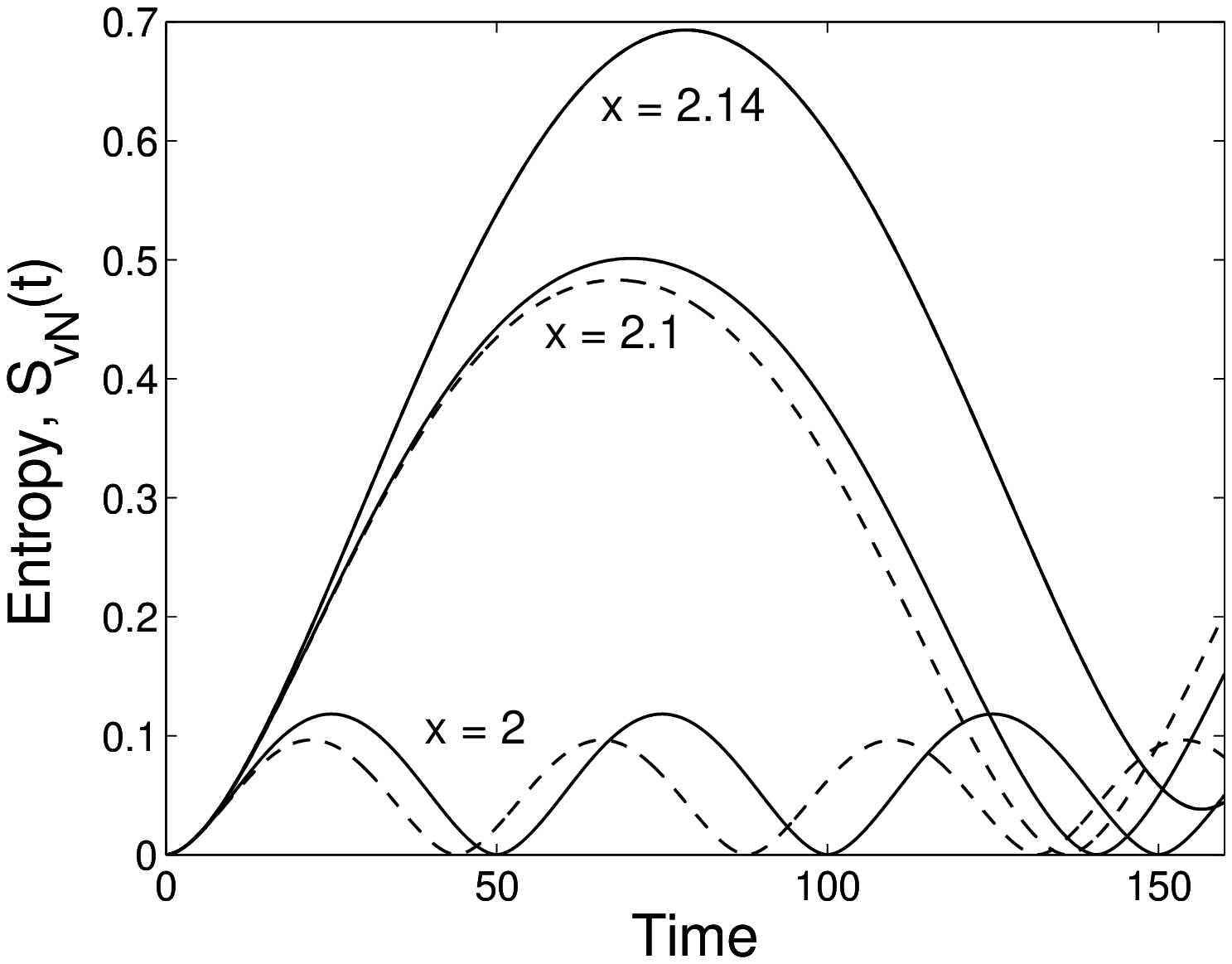}}
\caption{The time-evolution of the entropy $S_{\mathrm{vN}}(t)$ for
  the uncontrolled system of one qubit coupled to a one-particle
  environment, with $\gamma = 0.02$, $\omega_1 = 1$, and various
  values of $\omega_2$. Solid lines: $\omega_2 = (\pi - x)^{-1}$;
  dashed lines: $\omega_2 = \pi - x$ (with $x = 2, 2.1, 2.14$). The
  initial state is $|\Psi_0 \rangle$ of (\ref{initial}). For a given
  value of $\gamma$, closer frequencies $\omega_1$ and $\omega_2$
  enhance the interaction between the qubit and environment, causing
  stronger decoherence and longer revival times.}
\label{fig:deco}
\end{figure}

We set the unit of time, thereby introducing a natural system of units,
by arbitrarily choosing $\omega_1 = 1$ for all simulations (this implies
that one period of the first qubit's free evolution is $2 \pi$). Details
of the dynamics depend on the system parameters (i.e., the frequencies
and coupling constants for the uncontrolled system). In the simplest
case of the uncontrolled system of one qubit coupled to a one-particle
environment ($m = n = 1$), the initial state is $|\Psi_0 \rangle =
|-\rangle_1 \otimes |+\rangle_2$, and the time evolution can be solved
analytically:
\begin{eqnarray}
\fl |\Psi(t) \rangle = \rme^{-\rmi \gamma t/4} \left\{
\cos\left( \Omega t \right) |\!-\!+\rangle
+ \rmi \sin\left( \Omega t \right) \left[
\frac{\omega_1 - \omega_2}{2 \Omega} |\!-\!+\rangle
+ \frac{\gamma}{2 \Omega} |\!+\!-\rangle \right] \right\} , \\
\fl \rho_{\mathrm{q}}(t) 
= \cos^2 \left( \Omega t \right) |- \rangle\langle -|
+ \sin^2 \left( \Omega t \right) \left[
\frac{(\omega_1 - \omega_2)^2}{4 \Omega^2} |- \rangle\langle -|
+ \frac{\gamma^2}{4 \Omega^2} |+ \rangle\langle +| \right] ,
\end{eqnarray}
where we use a simplified notation: $|\!-\!+\rangle = |-\rangle_1
\otimes |+\rangle_2$, $|\!+\!-\rangle = |+\rangle_1 \otimes
|-\rangle_2$, and $\Omega = \frac{1}{2} [(\omega_1 - \omega_2)^2 +
\gamma^2]^{1/2}$ is the Rabi frequency. Due to discreteness of the
environment's spectrum, the loss of coherence is reversible. If the
transition frequencies are degenerate, $\omega_1 = \omega_2$, then the
state of the composite system, $|\Psi(t) \rangle$, oscillates between
two direct-product states, $|\!-\!+\rangle$ and $|\!+\!-\rangle$. In
this case, complete coherence revivals will occur whenever $\sin(\Omega
t) = 0$ or $\cos(\Omega t) = 0$, i.e., at times $t_k^{(\mathrm{deg})} =
k \pi/(2 \Omega)$ ($k \in \mathbb{N}$).  However, if $\omega_1 \neq
\omega_2$, then $|\Psi(t) \rangle$ oscillates between the initial
direct-product state $|\!-\!+\rangle$ and an entangled state (a
superposition of $|\!-\!+\rangle$ and $|\!+\!-\rangle$). Therefore,
complete coherence revivals will occur only when $\sin(\Omega t) = 0$,
i.e., at times $t_k = k \pi/\Omega$ ($k \in \mathbb{N}$). If $|\omega_1
- \omega_2| \ll \gamma$, then, in addition to the complete revivals at
times $t_k$, partial revivals will occur at times $t_k^{(\mathrm{part})}
\approx (k-\frac{1}{2}) \pi/\Omega$ ($k \in \mathbb{N}$). The maximum
loss of coherence depends on the values of $\gamma$ and $|\omega_1 -
\omega_2|$. For a given value of $\gamma$, closer frequencies enhance
the interaction between the qubit and environment, causing higher peak
values of decoherence (i.e., the entropy) and longer revival times.
Figure~\ref{fig:deco} shows the time-evolution of the entropy for the
uncontrolled system of one qubit and a one-particle environment, with
$\gamma = 0.02$, $\omega_1 = 1$, and various values of $\omega_2$. The
entropy dynamics shown in figure~\ref{fig:deco}, obtained by numerically
propagating the Schr\"{o}dinger equation (\ref{Schro}), and are in full
agreement with the analytical results above. In particular, we find the
first-revival times $t_1 \approx \{ 50.0, 140.7, 313.2 \}$ for $\omega_2
= (\pi - x)^{-1}$ and $t_1 \approx \{ 43.9, 136.1, 313.2 \}$ for
$\omega_2 = \pi - x$ with $x = \{ 2, 2.1, 2.14 \}$, respectively. These
values fully agree with the analytical formula for $t_k$ obtained
above. Also, for $x = 2.14$, the frequency difference $|\omega_1 -
\omega_2| \approx 0.00159$ is about one order of magnitude smaller than
$\gamma$, and, correspondingly, a partial revival is found numerically
at $t_1^{(\mathrm{part})} \approx 156.6$, in agreement with the
analytical result.

For the optimal control simulations below, the system parameters are
chosen to ensure complex dynamics and strong decoherence: values of
$\gamma/\omega$ are up to 0.02, which is significant for QC
applications, and the frequencies $\omega_i$ are close (but not equal),
to enhance the interaction. For one qubit coupled to a one-particle
environment $(m = n = 1)$, we choose
\begin{equation}
\omega_1 = 1, \ \ \ \omega_2 = ( \pi - 2.14 )^{-1} \approx 0.99841.
\label{omega_12}
\end{equation}
Imposing upper limits on the gate duration $( t_{\mathrm{f}} \leq 60 )$
and coupling constant $( \gamma \leq 0.02 )$ places the dynamics of the
uncontrolled system in the regime where decoherence increases
monotonically with time (before the entropy reaches its maximum value of
$S_{\mathrm{vN}} \approx \ln 2$). This dynamical regime approximates
some of the effects that the QIP would experience from a larger
environment, in particular, preventing restoration of coherence to the
qubit by uncontrolled revivals. Thus, any increase in coherence may be
attributed exclusively to the action of the control field.

\begin{figure}
\epsfxsize=0.7\textwidth \centerline{\epsffile{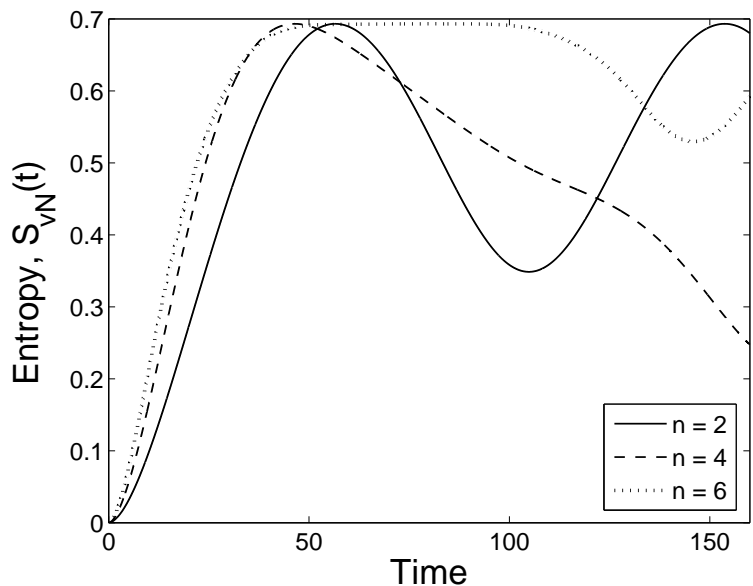}}
\caption{The time-evolution of the entropy $S_{\mathrm{vN}}(t)$ for
  the uncontrolled systems of one qubit coupled to $n$-particle
  environments: $n = 2$ (solid line), $n = 4$ (dashed line), and $n = 6$
  (dotted line). The coupling constant is $\gamma = 0.02$. Frequencies
  of the qubit, $\omega_1$, and the environmental particles, $\omega_j$
  ($j = 2, \ldots, n + 1$), are given by (\ref{omega_1j}). The initial
  state is $|\Psi_0 \rangle$ of (\ref{initial}).}
\label{fig:deco-n}
\end{figure}

When selecting the parameters of a multiparticle environment, we apply
the same criteria for maximizing decoherence of the uncontrolled system,
as described above. Figure~\ref{fig:deco-n} illustrates the uncontrolled
time-evolution of the entropy for a one qubit coupled to $n$-particle
environments ($n = 2, 4, 6$), with $\gamma = 0.02$. The frequencies of
the qubit and pairs of the environmental particles are given by
\begin{subequations}
\label{omega_1j} 
\begin{eqnarray}
& & \omega_1 = 1, \\
& & \omega_j = ( \pi - x_j )^{-1}, \ \ \ \omega_{j+1} = \pi - x_j,
\ \ \ j = 2, 4, \ldots, n, \\
& & x_j = \left\{ \begin{array}{ll} 2.14, & n = 2, \\ 
2.14, 2.1, & n = 4, \\ 
2.14, 2.1, 2, & n = 6. \end{array} \right.
\end{eqnarray}
\end{subequations}
For example, for $n = 4$, the frequencies of the four environmental
particles are approximately $\{ 0.96007, 0.99841, 1.00159, 1.04159 \}$.


\section{Optimal control algorithms}
\label{sec:optimize}

In the context of optimal control, the objective is to maximize the
fidelity of the target quantum gate over a set of time-dependent control
fields. The target quantum gates considered in this paper include the
Hadamard $( H_{\mathrm{t}} )$, identity $( I_2 )$, phase $( \pi/8 )$,
and controlled-NOT (CNOT) transformations:
\begin{subequations}
\begin{gather}
H_{\mathrm{t}} = \frac{1}{\sqrt{2}} \left( \begin{array}{cr} 
1 & 1 \\ 
1 & -1 \end{array} \right), \ \ \ 
I_2 = \left( \begin{array}{cr} 
1 & 0 \\
0 & 1 \end{array} \right), \ \ \ 
\frac{\pi}{8} = \left( \begin{array}{cc} 
1 & 0 \\ 
0 & \exp( \rmi \pi/4 ) \end{array} \right), 
\label{one} \\
\mathrm{CNOT} = \left( \begin{array}{cccc} 
1 & 0 & 0 & 0 \\ 
0 & 1 & 0 & 0 \\ 
0 & 0 & 0 & 1 \\ 
0 & 0 & 1 & 0 \end{array} \right). 
\label{CNOT}
\end{gather}
\end{subequations}
Collectively, $H_{\mathrm{t}}$, $\pi/8$, and CNOT constitute a universal
set of quantum gates for QC \cite{NC00}. Identity is included to
preserve an arbitrary quantum state during a specified time interval,
e.g., while operations are performed on other qubits.

In maximizing the gate fidelity, we employ a combination of two
optimization techniques, a genetic algorithm and a gradient
algorithm. For a given target gate, the genetic algorithm first locates
a parameterized control field that achieves a reasonable value of
fidelity (e.g., $F > 0.95$), then the gradient algorithm further
improves this result by lifting the parameterization restriction on the
field. This section describes the details of these search algorithms.

\subsection{Optimization with the genetic algorithm}

When the genetic algorithm is used, the gate fidelity $F$ is maximized
with respect to a parameterized control field
\begin{equation}
C(t) = f(t) \sum_{i=1}^{m} A_i \cos(\tilde{\omega}_i t + \theta_i ) ,  \
\ \ 0 \leq t \leq t_{\mathrm{f}} .
\label{gefield}
\end{equation}
Here, $f(t)$ is an envelope function incorporating the field's spectral
width, $t_{\mathrm{f}}$ is the gate duration, and $A_i$,
$\tilde{\omega}_i$, and $\theta_i$ are the amplitude, central angular
frequency, and relative phase of the $i$th component of the field,
respectively. A combination of these optimization parameters (called
``genes'') represents an ``individual'' whose ``fitness'' is defined as
the fidelity of the gate generated by the corresponding field. A
collection of individuals constitutes a ``population'' (we use
population sizes of $\sim 250$). At each generation, we evaluate the
fitness of all population members and create the next generation by
crossover and mutation of genes of the fittest individuals (crossover
and mutation rates are between 20 and 40 percent). A novelty of this
algorithm implementation is the inclusion of the control duration
$t_{\mathrm{f}}$ as one of the optimization parameters.

\subsection{Optimization with the gradient algorithm}
\label{sec:grad}

Removing the constraints on the control field imposed by the
parameterized form (\ref{gefield}) provides the potential for more
effective control of the system. In this case optimal control fields
are found by minimizing the following functional \cite{PaKo}:
\begin{equation}
K = J + \mathrm{Re} \int_{0}^{t_{\mathrm{f}}} \Tr \left\{ \left[
\dot{U}(t) + \rmi H(t) U(t) \right] B(t) \right\} \rmd t +
\frac{\alpha}{2} \int_{0}^{t_{\mathrm{f}}} \left| C(t) \right|^2 \rmd t.
\label{K-functional}
\end{equation}
In addition to the distance measure $J$ of (\ref{frobdist1}), $K$
includes a constraining term and a cost term. Upon minimization of $K$,
the first integral constrains $U(t)$ to obey the Schr\"{o}dinger
equation ($B(t)$ is an operator Lagrange multiplier) and the second
integral term penalizes the field fluence,
\begin{equation}
\mathcal{E} = \int_{0}^{t_{\mathrm{f}}} \left| C(t) \right|^2 \rmd t,
\label{fluence}
\end{equation}
with a weight $\alpha > 0$.

\subsubsection{Optimal control equations.}

An optimal control field is obtained by solving a set of equations that
follow from the variational analysis of $K$ as a functional of $B(t)$
and $U(t)$. Here, we derive the corresponding functional derivatives
\cite{RS80} and boundary conditions. The functional derivative of $K$
with respect to $B(t)$ yields
\begin{equation}
\frac{\delta K}{\delta B(t)} = \mathrm{Re} \left\{ \left[ \dot{U}(t) +
\rmi H(t) U(t) \right]^T \right\},
\end{equation}
so that the condition $\delta K/\delta B(t) = 0$ results in the
Schr\"{o}dinger equation (\ref{Schro}) for $U(t)$. Next we compute the
functional derivative of $K$ with respect to $U(t)$:
\begin{equation}
\frac{\delta K}{\delta U(t)} = \mathrm{Re} \left\{ \frac{\delta
J}{\delta U(t)} + B^T (t_{\mathrm{f}}) \delta(t - t_{\mathrm{f}}) -
\left[ \dot{B}(t) - \rmi B(t) H(t) \right]^T \right\}.
\end{equation}
Since $J$ depends only on $U = U(t_{\mathrm{f}})$, we obtain $\delta
J/\delta U(t) = (\rmd J/\rmd U) \delta(t - t_{\mathrm{f}})$. Therefore,
the condition $\delta K/\delta U(t) = 0$ results in two equations:
\begin{gather}
\dot{B}(t) = \rmi B(t)H(t),
\label{SchroB} \\
B^T (t_{\mathrm{f}}) = - \frac{\rmd J}{\rmd U}.
\label{BBC1} 
\end{gather}
We will also use the functional derivative of $K$ with respect to
$C(t)$,
\begin{subequations}
\label{dKdC}
\begin{gather}
\frac{\delta K}{\delta C(t)} = \mathrm{Im} \left\{ \Tr \left[ \hat{\mu}
    U(t) B(t) \right] \right\} + \alpha C(t), \\
\hat{\mu} = \sum_i^m \mu_i S_{ix},
\end{gather}
\end{subequations}
to guide the gradient search, as described in
section~\ref{seq:numerical} below.

The initial condition for $U(t)$ is $U(0) = I_{2^N}$ and the final
condition for $B(t)$ is given by (\ref{BBC1}). In order to find the
explicit form of $\rmd J/\rmd U$, first consider a scalar function
$y(Z(x))$, where $Z$ is a matrix function of the scalar variable
$x$. Using the chain rule, we obtain
\begin{equation}
\frac{\rmd y}{\rmd x} = \sum_{\kappa , \kappa'} 
\frac{\rmd y}{\rmd Z_{\kappa \kappa'}} 
\frac{\rmd Z_{\kappa \kappa'}}{\rmd x} 
= \sum_{\kappa , \kappa'} \frac{\rmd y}{\rmd Z_{\kappa \kappa'}} 
\frac{\rmd Z^{T}_{\kappa' \kappa}}{\rmd x}
= \Tr \left( \frac{\rmd y}{\rmd Z} \frac{\rmd Z^T}{\rmd x} \right).
\label{chrule}
\end{equation}
Setting $y = \Tr ( Z )$, implies that
\begin{equation}
\frac{\rmd y}{\rmd Z} = I.
\label{tderiv}
\end{equation}
Now let $Z = \left( Q^{\dag}Q \right)^{1/2}$ and $x = U_{a b}$ (a
complex scalar variable). The matrix indices $a$ and $b$ range from $1$
to $2^N$. Note that $\left( Q^{\dag}Q \right)^{1/2}$ is not an analytic
function of $U_{a b}$, but it can be expressed as an analytic function
of $U_{a b}$ and $U_{a b}^{\ast}$. Therefore, a generalized complex
derivative \cite{RS80} is applied to calculate $\rmd Z/\rmd x$, so that
$U_{a b}^{\ast}$ and subsequently $Q^{\dag}$ are treated as constants
when differentiating $\left( Q^{\dag}Q \right)^{1/2}$ with respect to
$U_{a b}$. Thus we find that
\begin{equation}
\frac{\rmd Z}{\rmd x} = 
\frac{\rmd \left( Q^{\dag}Q \right)^{1/2}}{\rmd U_{a b} }
= \frac{1}{2} \left( Q^{\dag}Q \right)^{-1/2} Q^{\dag} 
\frac{\rmd Q}{\rmd U_{a b}}.
\label{gcderiv}
\end{equation}
By combining (\ref{chrule})-(\ref{gcderiv}), we obtain
\begin{equation}
\frac{\rmd y}{\rmd x} = \frac{\rmd }{\rmd U_{a b}} 
\Tr \left[ \left( Q^{\dag}Q \right)^{1/2} \right]
= \frac{1}{2} \Tr \left[ \left( Q^{\dag}Q
  \right)^{-1/2}Q^{\dag}\frac{\rmd Q}{\rmd U_{a b}} \right].
\label{dydx}
\end{equation}
With the above notation, $J = (1 - 2^{-N} y)^{1/2}$. Noting that $(\rmd
J/\rmd U)_{a b} = \rmd J/\rmd U_{a b}$ and using (\ref{dydx}), we
finally derive
\begin{subequations}
\label{funcD}
\begin{gather}
\left( \frac{\rmd J}{\rmd U} \right)_{a b} 
= -\frac{2^{-N}}{4} \left\{ 1 - 2^{-N} \Tr \left[ 
\left( Q^{\dag}Q \right)^{1/2} \right]
\right\}^{-1/2} \Tr \left[ \left( Q^{\dag}Q \right)^{-1/2}
Q^{\dag}\frac{\rmd Q}{\rmd U_{a b}} \right], \\
\frac{\rmd Q}{\rmd U_{a b}} 
= G_{\lceil a/2^n \rceil \, \lceil b/2^n \rceil }^{\ast} 
|a\!\!\!\!\mod{2^n}\rangle\langle b\!\!\!\!\mod{2^n}|.
\label{delta_Q}
\end{gather}
\end{subequations}
Equation (\ref{delta_Q}) is obtained from (\ref{Q-def}), using the fact
that $k\!\!\mod{k} = k$. In (\ref{delta_Q}), the states are elements of
the environment's orthonormal basis $\{ |\nu\rangle \}$, and $\lceil x
\rceil$ denotes the smallest integer greater than or equal to $x$. The
explicit form of the boundary condition for $B(t)$ is obtained by
substituting (\ref{funcD}) into (\ref{BBC1}).

\subsubsection{The numerical procedure.}
\label{seq:numerical}

Optimal control fields are found using an iterative gradient algorithm
described below. An initial guess for the control field is needed at the
first iteration. Typically, we use the output of the genetic algorithm
as the initial guess for faster convergence, although fields of the form
(\ref{gefield}) with a random choice of parameters can be used as
well. At each iteration, $U(t)$ is propagating forward in time with the
Schr\"{o}dinger equation (\ref{Schro}) and the initial condition $U(0) =
I_{2^N}$. The resulting matrix $U = U(t_{\mathrm{f}})$ is used to
determine the final condition (\ref{BBC1}) for $B(t_{\mathrm{f}})$. Then
$B(t)$ is propagated backward in time with the time-reversed
Schr\"{o}dinger equation (\ref{SchroB}). All propagations are performed
using a toolkit for computational efficiency \cite{YMR03}. The resulting
$U(t)$ and $B(t)$ are utilized to compute the functional derivative
$\delta K/\delta C(t)$ of (\ref{dKdC}), which then adjusts the control
field for the next iteration. The adjustment of the control field for
the $k$th iteration $( k \in \mathbb{N} )$ is given by
\begin{equation}
C^{(k)}(t) = C^{(k-1)}(t) -  \beta \sin^r \left( \pi t / t_{\mathrm{f}}
\right) \left. \frac{\delta K}{\delta C(t)} \right|_{C(t) =
C^{(k-1)}(t)},
\end{equation}
where $0 < \beta \leq 1$ and $\frac{1}{2} \leq r \leq 1$ are constants
used to modify the magnitude of the field adjustment. The multiplier
$\sin^r \left( \pi t / t_{\mathrm{f}} \right)$ ensures that the control
field $C(t)$ is nearly zero at the initial and final time, which is a
reasonable physical restriction on the field. This iteration routine
continues until we observe no further improvement in $K$, which
manifests the achievement of convergence.

Despite the lack of direct coupling of the control field to the
environment, it can be shown that the composite system described by
(\ref{Ham}) is completely controllable (up to a global phase), as
defined in \cite{RSDRP95}. However, the restrictions on the gate
duration and on the shape of the control field limit the achievable
fidelity.


\section{Results of optimal control in the presence of decoherence}
\label{sec:results}

\subsection{One qubit coupled to a one-particle environment}

\begin{figure}
\epsfxsize=0.8\textwidth \centerline{\epsffile{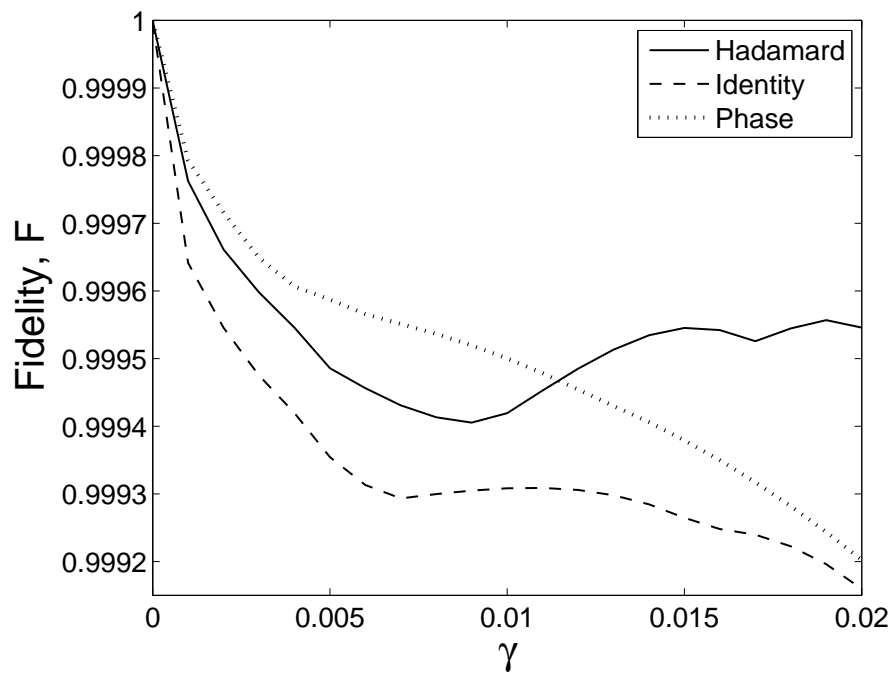}}
\caption{The gate fidelity $F$ versus the coupling constant $\gamma$,
  for optimally controlled one-qubit gates: Hadamard (solid line),
  identity (dashed line), and phase (dotted line). Each one-qubit gate
  is coupled to a one-particle environment. Values of $\gamma$ range
  from 0 to 0.02 in increments of 0.001.}
\label{fig:fit}
\end{figure}

We consider the optimally controlled Hadamard, identity, and phase gates
generated for a single qubit coupled to a one-particle environment $(m =
n = 1)$. Fidelities for these one-qubit gates are presented in
figure~\ref{fig:fit} for various values of the coupling constant
$\gamma$. The control fields optimized for the actual values of $\gamma$
result in fidelities above 0.9991. In particular, for the Hadamard
transform, we obtain $F > 1 - 10^{-6}$ for $\gamma = 0$ (a closed
system) and $F \approx 0.9995$ for $\gamma = 0.02$ (the strongest
coupling considered). In contrast, when the control field optimized for
$\gamma = 0$ is applied to the system with $\gamma = 0.02$, it generates
a gate with a poor fidelity, $F \approx 0.9063$. This result
demonstrates that optimal solutions designed for the ideal case of a
closed system have little value when applied to realistic open
systems. However, the optimal control algorithm is able to generate
quantum gates with very high fidelities, if coupling to the environment
is explicitly taken into account.

\begin{figure}
\epsfxsize=0.8\textwidth \centerline{\epsffile{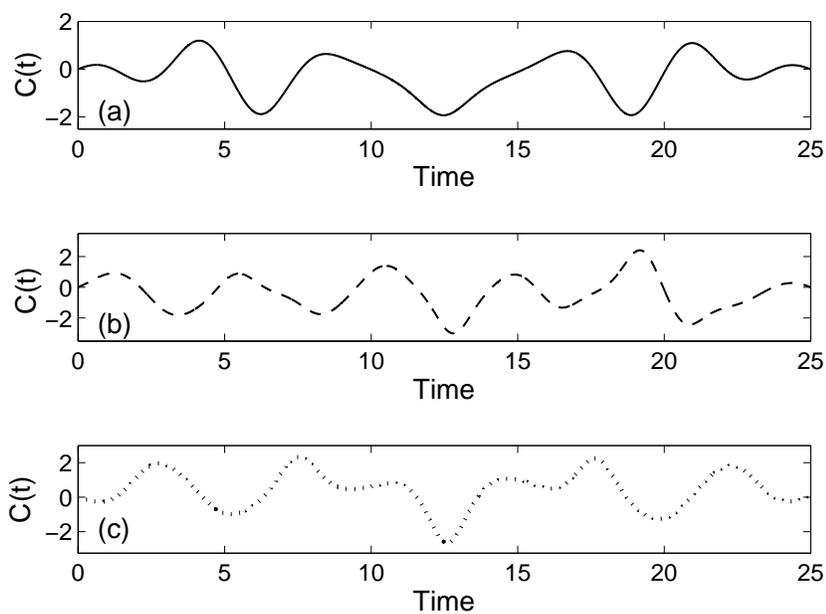}}
\caption{Optimal control fields $C(t)$ versus time, for one-qubit
  gates: (a) Hadamard, (b) identity, and (c) phase. Each one-qubit gate
  is coupled to a one-particle environment ($\gamma = 0.02$).}
\label{fig:field}
\end{figure}

The optimal control fields that generate the one-qubit gates (with a
one-particle environment and $\gamma = 0.02$) are shown in
figure~\ref{fig:field}. These fields are intense, with maximum
amplitudes larger than $2.0$ (in the units of $\hbar = \omega_1 = \mu_i
= 1$). The gate duration is $t_{\mathrm{f}} = 25.0$ (about four periods
of free evolution). The exact time structure of an optimal field is not
intuitive and is tailored to the particular control application. For
example, control fields optimized for $\gamma = 0.02$ are not only more
intense than those optimized for $\gamma = 0$, they also have very
different structures. One common feature of the optimal control fields
presented in figure~\ref{fig:field} is that they are approximately
symmetric about $t \approx t_{\mathrm{f}} /2$. We suggest that this
property of the fields is related to the reversibility of the system
dynamics: the periods in which the information flows from the QIP to the
environment are followed by periods in which the information flow is
reversed, in order to restore the coherence of the QIP.

\begin{figure}
\epsfxsize=0.8\textwidth \centerline{\epsffile{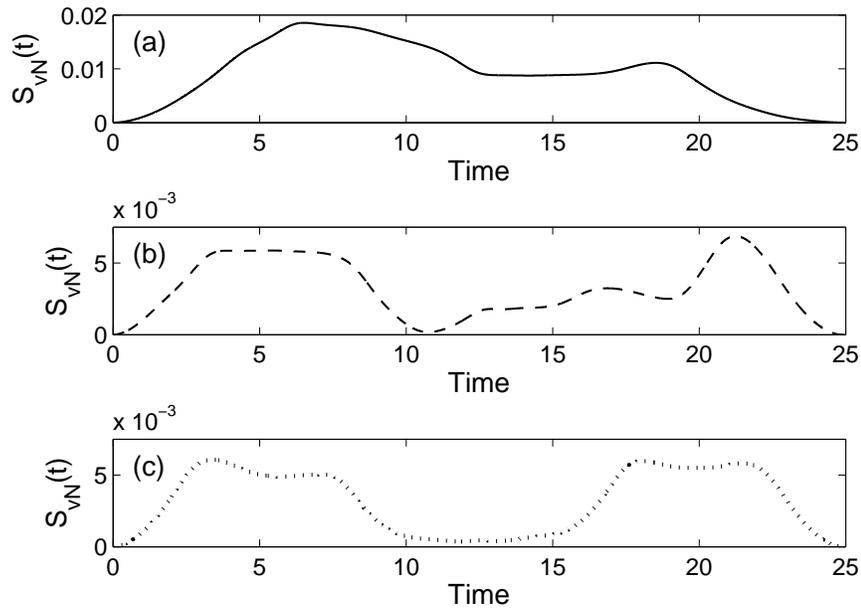}}
\caption{The von Neumann entropy $S_{\mathrm{vN}}(t)$ versus time, for
  optimally controlled one-qubit gates: (a) Hadamard, (b) identity, and
  (c) phase. Each one-qubit gate is coupled to a one-particle
  environment ($\gamma = 0.02$). The initial state is $|\Psi_0 \rangle$
  of (\ref{initial}).}
\label{fig:fit-deco}
\end{figure}

Figure~\ref{fig:fit-deco} shows the time behavior of the von Neumann
entropy of the QIP for optimally controlled one-qubit gates (with
$t_{\mathrm{f}} = 25.0$ and $\gamma = 0.02$). By comparing
figures~\ref{fig:fit-deco}~and~\ref{fig:deco}, we observe that the
optimal control dramatically enhances coherence of the qubit system in
comparison to the uncontrolled dynamics. Decoherence is suppressed by
the control at all times, but especially at the end of the gate
operation (i.e., for $t = t_{\mathrm{f}}$). For example,
$S_{\mathrm{vN}}(t_{\mathrm{f}}) < 10^{-7}$ for the Hadamard gate with
$\gamma = 0.02$, which means that at $t = t_{\mathrm{f}}$ the qubit
system and environment are almost completely uncoupled. Inspecting
eigenvalues of the controlled Hamiltonian, we find that the intense
control field creates significant dynamic shifts of the energy
levels. Specifically, under the influence of the optimal control
field, four of the six transition frequencies of the composite system
experience high-amplitude oscillations (following the corresponding
changes in the field strength). This effect is mainly responsible for
reducing the qubit-environment interaction during the control
pulse. However, achieving extremely low final-time entropies and
correspondingly high gate fidelities requires the employment of an
induced coherence revival. For the selected set of the system
parameters, revivals in the uncontrolled dynamics occur at times much
longer than $t_{\mathrm{f}}$ (specifically, $t_1^{(\mathrm{part})}
\approx 156.6$ and $t_1 \approx 313.2$), so that the almost complete
coherence revival observed at $t = t_{\mathrm{f}}$ is induced
exclusively by the control field.

For very short gate durations $( t_{\mathrm{f}} < 5 )$, a different type
of optimal solution is found. The control fails to induce revivals at
such short times and therefore generates gates with smaller fidelities
(e.g., $F \approx 0.9874$ for the Hadamard transform with $\gamma =
0.02$ and $t_{\mathrm{f}} \approx 2.33$). In this short-time regime the
control relies on the decoherence suppression via dynamic shifting of
the energy levels and on very fast operation (trying to perform the
target transformation in the shortest time possible to limit the effect
of decoherence), but not on the creation of coherence revivals. Such
short-time controls can be useful for environments with very dense
spectra, for which the induced-revival times will be impractically long.

We study in detail how the choice of the control duration
$t_{\mathrm{f}}$ affects properties of the optimal control field, gate
fidelity, and decoherence dynamics. Specifically, we optimize the
one-qubit Hadamard gate (with a one-particle environment and $\gamma =
0.02$) for all integer values of $t_{\mathrm{f}}$ between 2 and 40 using
the gradient algorithm described in section~\ref{sec:grad}. For
$t_{\mathrm{f}} < 5$ we find the fast-control no-revival regime
described above. Interestingly, most optimal control fields with
$t_{\mathrm{f}} > 5$, in addition to inducing an almost complete
coherence revival at the final time, also produce a partial revival at
approximately $t_{\mathrm{f}} /2$. Optimal control fields with $5 <
t_{\mathrm{f}} < 20$ typically exhibit large amplitudes and fluences and
strong low-frequency components. For $t_{\mathrm{f}} = 25$ we find the
optimal control field that generates the quantum gate with a better
fidelity while having a smaller amplitude and fluence, as compared to
the fields obtained for shorter control durations. As $t_{\mathrm{f}}$
increases to 25, the gate fidelity increases to approximately 0.9995,
the final-time entropy decreases to approximately $10^{-7}$, and the
maximum field amplitude decreases to approximately 2.0. However,
increasing $t_{\mathrm{f}}$ above $25$ does not improve the optimal gate
performance; the field amplitudes, gate fidelities, and final-time
entropy values change very slightly for $25 \leq t_{\mathrm{f}} \leq
40$. The physical interpretation of this behavior is that the control
requires some time ($t_{\mathrm{f}} \geq 25$ in the present case) to
almost completely reverse the information flow between the QIP and
environment, and induce a nearly perfect coherence revival. From these
results, it appears that the pulse duration is a very important
characteristic of the control fields employed for quantum gate
generation.

\subsection{The Kraus-map dynamics of the qubit}

The time-dependent state of the QIP, which is coupled to the
environment, is represented by the reduced density matrix
(\ref{red-dens}). In order to examine the reduced dynamics of the QIP,
it is instructive to use the Kraus-map representation \cite{Kraus}. If
the composite system was initially (i.e., at time $t = 0$) in the
direct-product state,
\begin{equation}
\rho(0) = \rho_{\mathrm{q}}(0) \otimes \rho_{\mathrm{env}}(0)
= \rho_{\mathrm{q}}(0) \otimes 
\sum_{\nu = 1}^{2^n} \varrho_{\nu} |\nu \rangle\langle \nu|,
\end{equation}
then the reduced dynamics of the QIP has the following form (known as
the Kraus map \cite{Kraus}):
\begin{equation}
\rho_{\mathrm{q}}(t) = \Phi[\rho_{\mathrm{q}}(0)]  = \sum_{\nu, \nu' =
1}^{2^n} K_{\nu \nu'}(t) \rho_{\mathrm{q}}(0) K_{\nu \nu'}^{\dag}(t),
\end{equation}
where the Kraus operators $K_{\nu \nu'}(t) \in M_{2^m}(\mathbb{C})$ are
given by
\begin{subequations}
\begin{gather}
K_{\nu \nu'}(t) = \sqrt{\varrho_{\nu'}} \sum_{i, i' = 1}^{2^m}
U_{\substack{i i' \\ \nu \nu'}}(t) |i\rangle\langle i'|, 
\label{K-operators} \\  
\sum_{\nu, \nu' = 1}^{2^n} K_{\nu \nu'}^{\dag}(t) K_{\nu \nu'}(t) =
I_{2^m}.
\end{gather}
\end{subequations}
It is well known \cite{Kraus} that there exist infinitely many different
sets of Kraus operators, $\{ K_1, \ldots, K_p \}$ (where $p \in
\mathbb{N}$ is the number of operators in the set), that represent the
same map $\Phi$ (i.e., they evolve $\rho_{\mathrm{q}}(0)$ in exactly the
same way). Moreover, any Kraus map for an $k$-level quantum system can
be represented by a set of $p \leq k^2$ Kraus operators. That is, if the
map is represented by a set of $p' > k^2$ Kraus operators, there always
exists another representation with not more than $k^2$
operators. Therefore, for our system of $m$ qubits and $n$ environmental
particles, the set of $2^{2 n}$ Kraus operators (\ref{K-operators}) can
always be transformed into another set of not more than $2^{2 m}$
operators, representing the same map $\Phi$.  However, since we
numerically study Kraus operators only for the case of $n = m = 1$,
there is no practical need for such a transformation.

\begin{figure}
\epsfxsize=0.8\textwidth \centerline{\epsffile{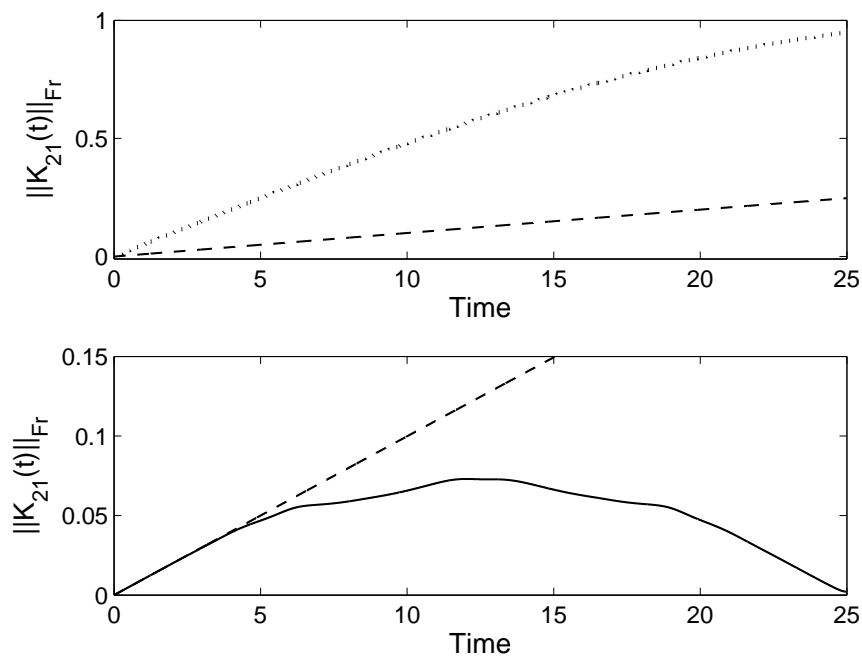}}
\caption{The time-evolution of the Kraus operator's norm, $\|
  K_{21}(t) \|_{\mathrm{Fr}}$, for the system of one qubit and one
  environmental particle: uncontrolled evolution with $\gamma = 0.1$
  (dotted line), uncontrolled evolution with $\gamma = 0.02$ (dashed
  lines), and controlled evolution, under the optimal control field
  generating the Hadamard gate, with $\gamma = 0.02$ (solid line). The
  initial state is $|\Psi_0 \rangle$ of (\ref{initial}).}
\label{fig:K12}
\end{figure}

In calculations, we use $\rho(0) = |\Psi_0 \rangle \langle \Psi_0 |$
with $|\Psi_0 \rangle$ of (\ref{initial}). For one qubit coupled to a
one-particle environment, we use the notation $|\nu = 1\rangle =
|+\rangle$ and $|\nu = 2\rangle = |-\rangle$ and find $K_{12}(t) =
K_{22}(t) = 0$ and $K_{11}^{\dag}(t) K_{11}(t) + K_{21}^{\dag}(t)
K_{21}(t) = I_2$. It is therefore sufficient to explore either
$K_{11}(t)$ or $K_{21}(t)$.  By evaluating the Kraus operators we can
quantify the non-unitarity of the qubit dynamics. It is important to
note that the non-unitary evolution is not only responsible for
decoherence, but is also required to steer the information flow back to
the QIP. The control field that restores coherence to the QIP
necessarily employs the interaction with the environment and the
corresponding non-unitary dynamics. We examine the time behavior of the
Frobenius norm of the Kraus operator, $\| K_{21}(t) \|_{\mathrm{Fr}}$,
that serves as a measure of non-unitarity. Figure~\ref{fig:K12} shows
$\| K_{21}(t) \|_{\mathrm{Fr}}$ for both controlled and uncontrolled
dynamics. In comparison to the uncontrolled evolution, the optimal
control dramatically decreases the non-unitarity of the qubit dynamics
during the gate operation, culminating in almost complete decoupling at
the final time $t_{\mathrm{f}}$. We also see that, under the optimal
control, $\| K_{21}(t) \|_{\mathrm{Fr}}$ is approximately symmetric
about $t \approx t_{\mathrm{f}} /2$. Inspecting the time derivative of
the entropy, $\rmd S_{\mathrm{vN}}/\rmd t$, we find that $\| K_{21}(t)
\|_{\mathrm{Fr}}$ reaches the maximum at approximately the same time
(just prior to $t_{\mathrm{f}} /2$) when the fastest decrease in the
qubit's entropy is observed, indicating the maximum flow of information
into the QIP.

\subsection{One qubit coupled to a multiparticle environment}

We explore the performance of optimally controlled one-qubit gates in
the presence of multiparticle environments described in
section~\ref{sec:model}. Table~\ref{results} reports optimal control
field parameters, fidelity, and final-time entropy for the one-qubit
Hadamard gate coupled to $n$-particle environments $(m = 1$, $n = 1,
2, 4, 6$, and $\gamma = 0.02$). For $n = 4$, the values in
Table~\ref{results} were obtained with the coupling scheme modeled by
a two-dimensional lattice of (\ref{square_c}); however, very similar
results were obtained with the linear nearest-neighbor coupling scheme
of (\ref{nnc}).

The results obtained for $n \geq 2$ further illustrate the benefits of
optimal controls which explicitly take into account coupling to the
environment. The entropy dynamics indicate that for multiparticle
environments the control employs the same mechanism of an induced
coherence revival, as described above for $n = 1$. Fast and intense
control fields significantly suppress the qubit-environment interaction
during the gate operation and try to recover as much of the lost
information as possible before the end of the control pulse.  However,
as the complexity of the composite system increases, it becomes more
difficult to induce an almost perfect revival; therefore, the gate
fidelity and final-time coherence decrease as $n$ increases.  This
observation supports the conclusion that shorter-time controls (which do
not rely on revivals) will be useful for environments with dense spectra.

\begin{table}[t]
\caption{The performance of the optimally controlled one-qubit
  Hadamard gate in the presence of various $n$-particle environments
  ($\gamma = 0.02$). Here, $A_{\mathrm{max}}$, $t_{\mathrm{f}}$,
  $\mathcal{E}$, $F$, and $S_{\mathrm{vN}}(t_{\mathrm{f}})$ are the
  maximum field amplitude, control duration, field fluence,  gate
  fidelity, and final-time entropy, respectively. $F_{\gamma = 0}$
  denotes the gate fidelity obtained when the control field optimized
  for $\gamma = 0$ is applied to the system with $\gamma = 0.02$. The
  initial state for the entropy computation is $|\Psi_0 \rangle$ of
  (\ref{initial}).}
\label{results} 
\begin{indented}
\item[]\begin{tabular}{@{}ccccc}
\br
$n$ & 1 & 2 & 4 & 6 \\
\mr
$A_{\mathrm{max}}$ & 2.0 & 4.0 & 4.0 & 2.5 \\
$t_{\mathrm{f}}$ & 25.0 & 15.4 & 25.0 & 25.0 \\
$\mathcal{E}$ & 20.0 & 49.0 & 55.5 & 34.0 \\
\mr
$F$ & 0.9995 & 0.9975 & 0.9935 & 0.9786 \\
$F_{\gamma = 0}$ & 0.9063 & 0.8829 & 0.8133 & 0.7723 \\
$S_{\mathrm{vN}}(t_{\mathrm{f}})$ & $9.0 \times 10^{-8}$ & 
$4.4 \times 10^{-5}$ & $4.7 \times 10^{-4}$ & $3.0 \times 10^{-3}$ \\
\br
\end{tabular}
\end{indented}
\end{table}

\subsection{Two qubits with a one-particle environment}
\label{sec:CNOT}

\begin{figure}
\epsfxsize=0.8\textwidth \centerline{\epsffile{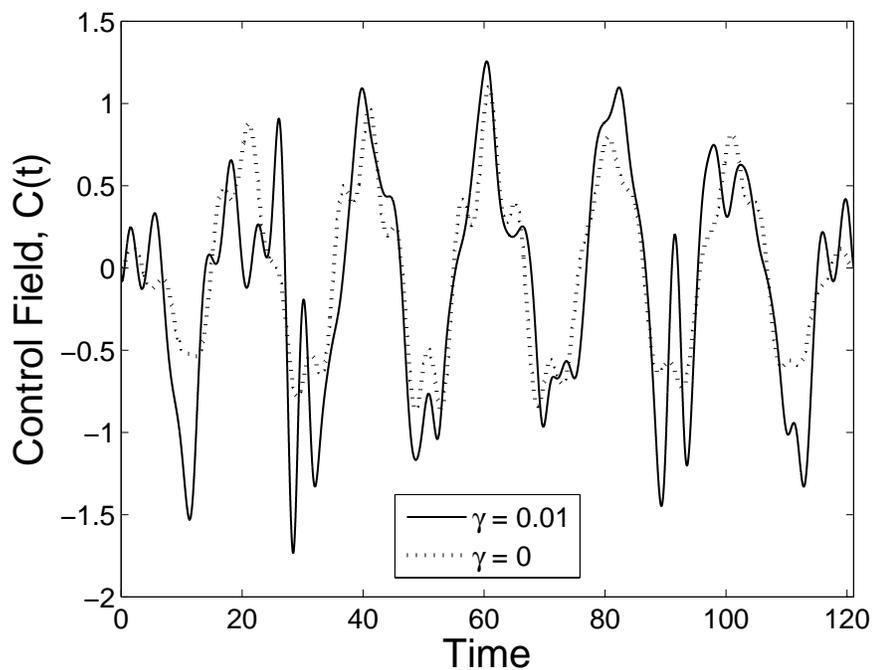}}
\caption{Optimal control fields $C(t)$ versus time, for the two-qubit
  CNOT gate with $\gamma = 0.01$ (solid line) and $\gamma = 0$ (dotted
  line).}
\label{fig:field-CN}
\end{figure}

\begin{figure}
\epsfxsize=0.8\textwidth \centerline{\epsffile{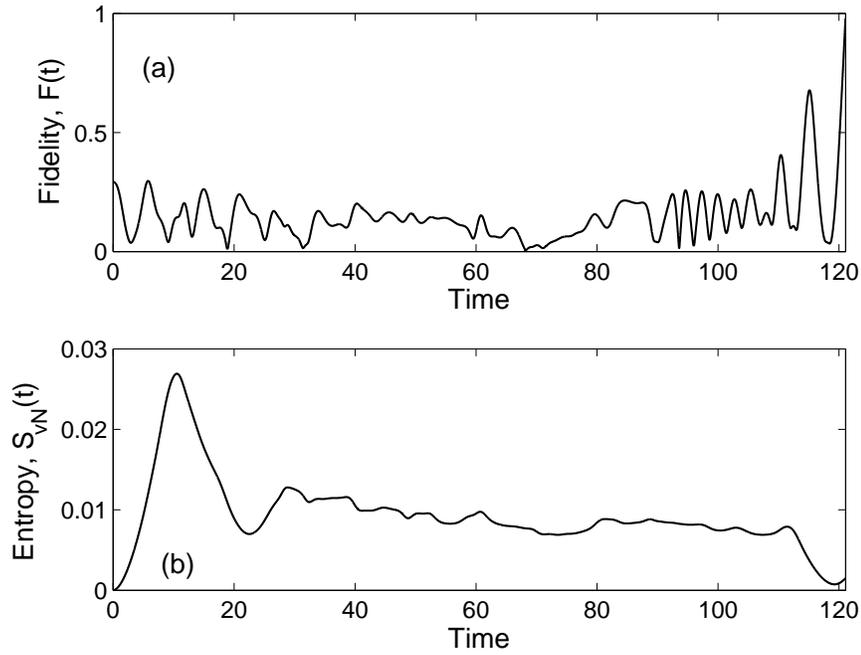}}
\caption{The time-evolution of (a) fidelity $F(t)$ and (b) von Neumann
  entropy $S_{\mathrm{vN}}(t)$ during the optimally controlled CNOT gate
  operation ($m = 2$, $n = 1$, and $\gamma = 0.01$). These results
  demonstrate that a high degree of coherence (quantified by the
  entropy) does not ensure a correspondingly high value of the gate
  fidelity. The initial state for the entropy computation is $|\Psi_0
  \rangle$ of (\ref{initial}).}
\label{fig:fid-entropy}
\end{figure}

For the QIP consisting of two qubits ($m = 2$), the target gate is CNOT
of (\ref{CNOT}). The coupling constant between the two qubits is
$\gamma_{12} = 0.1$, while the coupling constant between each qubit and
the single environmental particle ($n = 1$) is $\gamma_{13} =
\gamma_{23} = \gamma$. Frequencies of the two qubits are $\omega_1 = 1$
and $\omega_2 = \pi - 2.05 \approx 1.09519$, and the frequency of the
environmental particle is $\omega_3 = \left( \pi - 2.14 \right)^{-1}
\approx 0.99841$. The optimal control fields obtained for $\gamma = 0$
and $\gamma = 0.01$ (shown in figure~\ref{fig:field-CN}) generate the
CNOT gate with fidelities of 0.9999 and 0.9798, respectively. When
$\gamma = 0.01$, the entropy for the uncontrolled evolution increases
monotonically until $t \approx 125$ (reaching a maximum of approximately
0.6), whereas the optimal control field results in a much lower entropy,
shown in sub-plot (b) of figure~\ref{fig:fid-entropy}. The same pattern
of a partial revival at an intermediate time followed by an almost
complete revival at $t = t_{\mathrm{f}}$, seen for the one-qubit gates
in figure~\ref{fig:fit-deco}, is also present for the two-qubit gate,
but on a longer time scale. For the CNOT gate's final-time coherence
revival we find $S_{\mathrm{vN}}(t_{\mathrm{f}}) \approx 1.5 \times
10^{-3}$ at $t_{\mathrm{f}} = 121.1$.

We observe that the fidelity of the optimally controlled quantum gates
decreases with increases in $n$ (the number of environmental particles)
and, even more significantly, $m$ (the number of qubits in the
QIP). This behavior arises due to the difference between the perfect
control solution and an actual control field found by the optimization
algorithm. According to an analysis of the control landscape for unitary
transformations \cite{RHmR05, BRHWK06}, the pernicious effect of control
inaccuracies on the gate fidelity rapidly increases with the size of the
system. If instead of the perfect control solution $C_0 (t)$, the actual
field is $C_0 (t) + \delta C(t)$, then instead of the perfect fidelity
$F = 1$, one will obtain $F = 1 - \delta F$, where $\delta F \propto 2^m
\| \delta C(t) \|^2$ (here, $\| \cdot \|$ denotes an appropriate
functional norm). As the number of interacting qubits, $m$, increases,
the factor $2^m$ becomes more important. Moreover, as the complexity of
the composite system increases (more qubits and/or environmental
particles), the control error $\| \delta C(t) \|$ will increase as well,
as it will become more difficult to find a field that is very close to
the perfect one.

\subsection{Can the state purity measure the gate fidelity?}

We found that obtaining a very high gate fidelity requires an almost
complete coherence revival characterized by a very low final-time
entropy. Is it then possible to rely on a characteristic of coherence
(e.g., the final-time entropy or purity of the QIP state) as a measure
of the gate quality, instead of measuring the distance between the
actual and target gate transformations? The answer is definitely ``no''
because the restoration of coherence is a necessary, but not sufficient,
condition for a high gate fidelity. There exist an infinite number of
unitary, or almost unitary, transformations which nevertheless are very
far from the target one.

In order to further emphasize this point, we generalize the notion of
the gate fidelity (as measured by the distance between the actual
evolution operator $U(t)$ and target transformation $G$) to all times $0
\leq t \leq t_{\mathrm{f}}$. Figure~\ref{fig:fid-entropy} shows this
time-dependent fidelity $F(t)$ and the entropy $S_{\mathrm{vN}}(t)$ for
the optimally controlled two-qubit CNOT gate (with $\gamma = 0.01$ and
$t_{\mathrm{f}} = 121.1$). We see that the minimum of the entropy occurs
at a time $t_{S_\mathrm{min}} \approx 119$ (i.e., before
$t_{\mathrm{f}}$) when $F(t)$ is still quite low, and that at the time
interval between $t_{S_\mathrm{min}}$ and $t_{\mathrm{f}}$, while the
fidelity $F(t)$ rapidly increases to achieve its final-time value $F
\approx 0.9798$, the entropy slightly increases as well. This example
shows that fidelity and coherence do not always correlate and that a
very low value of the entropy does not always result in a
correspondingly high value of the gate fidelity. According to this
analysis, a strategy of maximizing the state purity \cite{GK05, Hoh06}
does not ensure the generation of target quantum gates with the highest
possible fidelity.


\section{Robustness of optimally controlled gates to system
  variations}
\label{sec:robust}

We observed that applying the control field optimized for the closed
system ($\gamma = 0$) to the coupled one ($\gamma = 0.02$) results in a
significant decrease in the gate fidelity. Analogously, we find that
applying the control field optimized for the case of a one-particle
environment ($n = 1$) to systems with $n \geq 2$ environmental particles
also has a strong detrimental effect on the gate fidelity. These
results are part of a broader analysis of the robustness of optimally
controlled quantum gates to different types of system variations.

We address some aspects of this issue by considering the one-qubit
Hadamard gate, with a fixed number $n$ of environmental particles ($n =
1, 2, 4$), and finding an optimal control field for a specified set of
system parameters: the coupling constants $\gamma_{i j}$ given by
(\ref{gamma}) (with $\gamma = 0.02$) and frequencies $\omega_i$ given by
(\ref{omega_12}) for $n = 1$ and (\ref{omega_1j}) for $n \geq 2$. Then
we apply this control field to an ensemble of systems with normal
variations in either coupling constants $\gamma_{i j}$ or frequencies
$\omega_i$ and analyze how the uncertainties in the system parameters
affect the gate fidelity $F$ and final-time entropy
$S_{\mathrm{vN}}(t_{\mathrm{f}})$. Although the dependence of $F$ and
$S_{\mathrm{vN}}(t_{\mathrm{f}})$ on the coupling constants and
frequencies is highly non-linear (which implies that the distributions
of $F$ and $S_{\mathrm{vN}}(t_{\mathrm{f}})$ will not be normal), our
statistical analysis employs only mean values and standard deviations,
given by $\overline{F} = L^{-1} \sum_{r = 1}^{L} F_r$ and $\sigma_F = [
L^{-1} \sum_{r = 1}^{L} ( F_r - \overline{F} )^2 ]^{1/2}$, respectively,
for the gate fidelity $F$, and similarly for the final-time entropy
$S_{\mathrm{vN}}(t_{\mathrm{f}})$. The summation is over all elements of
the ensemble (ensemble sizes $L$ of the order of $10^5$ are used in the
calculations).

\subsection{Variation of the coupling constants}

\begin{figure}
\epsfxsize=0.8\textwidth \centerline{\epsffile{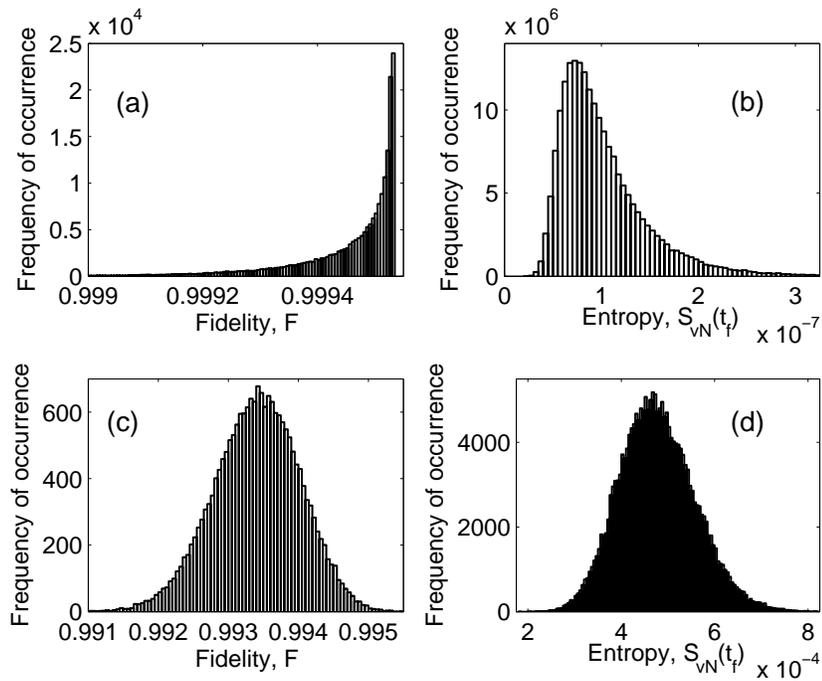}}
\caption{Frequency histograms for the gate fidelity and final-time
  entropy distributions, obtained when the control field optimized for
  the Hadamard gate with $\gamma = 0.02$ is applied to an ensemble of
  systems with normal variations in the coupling constants $\gamma_{i
  j}$. The distribution for each non-zero $\gamma_{i j}$ is normal with
  a mean $\overline{\gamma} = 0.02$ and a standard deviation
  $\sigma_{\gamma} = \overline{\gamma}/8 = 0.0025$. Sub-plots include
  frequency histograms of (a) the fidelity distribution for $n = 1$, (b)
  the entropy distribution for $n = 1$, (c) the fidelity distribution
  for $n = 4$, and (d) the entropy distribution for $n = 4$. Note the
  axes scale differences in the sub-plots. Table~\ref{tab:robust}
  reports statistical data for these distributions.}
 \label{fig:RFO-c}
\end{figure}

The value of each non-zero coupling constant $\gamma_{i j}$ (given by
(\ref{gamma}) with $\gamma = 0.02$) is individually replaced by a value
randomly selected from a normal distribution with a mean
$\overline{\gamma} = 0.02$ and a standard deviation $\sigma_{\gamma} =
\overline{\gamma}/8 = 0.0025$. The statistical analysis of the
corresponding distributions of the fidelity and final-time entropy is
reported in table~\ref{tab:robust}, and frequency histograms of these
distributions are shown in figure~\ref{fig:RFO-c}. These results
demonstrate a high degree of robustness of the performance of the
optimally controlled gate to relatively large variations in the strength
of the system-environment coupling. On average, there is practically no
decrease in the fidelity and entropy, and the relative width of the
fidelity distribution, $\sigma_F / \overline{F}$, is by several orders
of magnitude smaller than $\sigma_{\gamma} /
\overline{\gamma}$. Interestingly, if the control field optimized for
$\gamma = 0.02$ is applied to the closed system with $\gamma = 0$, this
results in a relatively high fidelity (e.g., $F = 0.9989$ for $n =
1$). The standard deviation $\sigma_F$ rises with the increase in the
number of environmental particles. We also see that the distributions of
$F$ and $S_{\mathrm{vN}}(t_{\mathrm{f}})$ are more symmetric for $n = 4$
than for $n = 1$.

\begin{table}
\caption{Fidelity and entropy data for the one-qubit Hadamard gate
  applied to an ensemble of systems with normal variations in the
  coupling constants $\gamma_{i j}$ and frequencies $\omega_i$.  Columns
  of $F$ and $S_{\mathrm{vN}}(t_{\mathrm{f}})$ contain fidelity and
  final-time entropy values, respectively, for the original system
  parameters: $\gamma = 0.02$ and frequencies given by (\ref{omega_12})
  for $n = 1$ and (\ref{omega_1j}) for $n \geq 2$. Columns of
  $\overline{F}$ and $\overline{S_{\mathrm{vN}}}$ contain mean values of
  fidelity and final-time entropy, respectively, over the ensemble,
  while $\sigma_F$ and $\sigma_{S_{\mathrm{vN}}}$ are the respective
  standard deviations.}
\label{tab:robust}
\begin{indented}
\lineup
\item[]\begin{tabular}{@{}ccccccc}
\br
\centre{7}{Variation in $\gamma_{i j}$} \\
\ms
$n$ & $F$ & $\overline{F}$ & $\sigma_F$ &
$S_{\mathrm{vN}}(t_{\mathrm{f}})$ & $\overline{S_{\mathrm{vN}}}$ &
$\sigma_{S_{\mathrm{vN}}}$ \\
\mr
1 & 0.9995 & 0.9995 & $1.1 \times 10^{-4}$ & $9.0 \times 10^{-8}$ & $1.0
\times 10^{-7}$ & $4.7 \times 10^{-8}$ \\
2 & 0.9975 & 0.9975 & $2.6 \times 10^{-4}$ & $4.4 \times 10^{-5}$ & $4.6
\times 10^{-5}$ & $1.5 \times 10^{-5}$ \\
4 & 0.9935 & 0.9934 & $6.1 \times 10^{-4}$ & $4.7 \times 10^{-4}$ & $4.8
\times 10^{-4}$ & $8.1 \times 10^{-5}$ \\
\mr
\centre{7}{Variation in $\omega_i$} \\
\ms
$n$ & $F$ & $\overline{F}$ & $\sigma_F$ &
$S_{\mathrm{vN}}(t_{\mathrm{f}})$ & $\overline{S_{\mathrm{vN}}}$ &
$\sigma_{S_{\mathrm{vN}}}$ \\
\mr
1 & 0.9995 & 0.9821 & $1.1 \times 10^{-2}$ & $9.0 \times 10^{-8}$ & $6.8
\times 10^{-3}$ & $7.4 \times 10^{-3}$ \\
2 & 0.9975 & 0.9896 & $5.3 \times 10^{-3}$ & $4.4 \times 10^{-5}$ & $7.0
\times 10^{-4}$ & $6.2 \times 10^{-4}$ \\
4 & 0.9935 & 0.9884 & $4.5 \times 10^{-3}$ & $4.7 \times 10^{-4}$ & $1.7
\times 10^{-3}$ & $1.8 \times 10^{-3}$ \\
\br
\end{tabular}
\end{indented}
\end{table}

\subsection{Variation of the frequencies}

\begin{figure}
\epsfxsize=0.8\textwidth \centerline{\epsffile{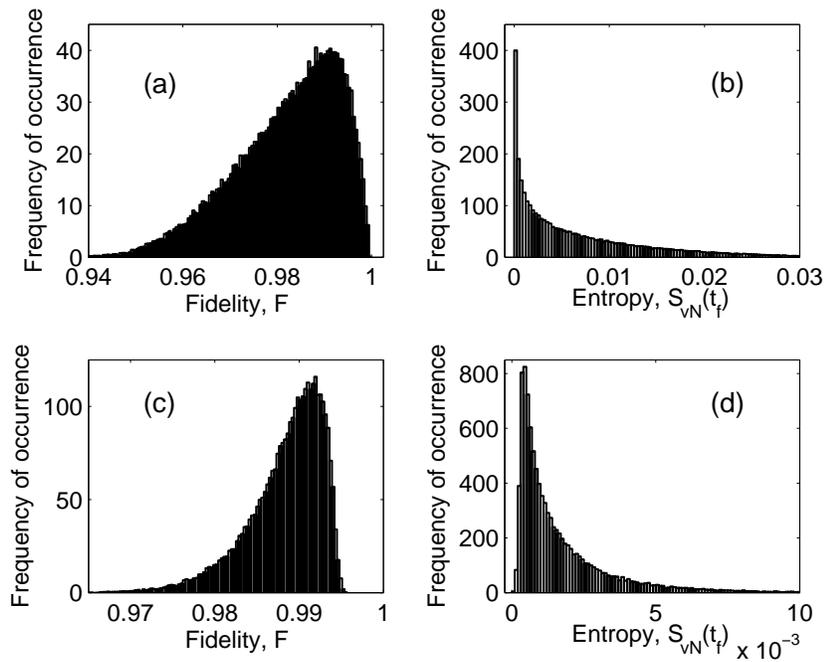}}
\caption{Frequency histograms for the gate fidelity and final-time
  entropy distributions, obtained when the control field optimized for
  the Hadamard gate with transition frequencies $\omega_i$ given by
  (\ref{omega_12}) for $n = 1$ and (\ref{omega_1j}) for $n \geq 2$ is
  applied to an ensemble of systems with normal variations in the
  transition frequencies. The distribution for each transition frequency
  is normal with a mean $\overline{\omega}_i = \omega_i$ and a standard
  deviation $\sigma_{\omega_i} = \omega_i /25$. Sub-plots include
  histograms of (a) the fidelity distribution for $n = 1$, (b) the
  entropy distribution for $n = 1$, (c) the fidelity distribution for $n
  = 4$, and (d) the entropy distribution for $n = 4$. Note the axes
  scale differences in the sub-plots. Table~\ref{tab:robust} reports
  statistical data for these distributions.}
  \label{fig:RFO-f}
\end{figure}

The value of each frequency $\omega_i$ (given by (\ref{omega_12}) for $n
= 1$ and (\ref{omega_1j}) for $n \geq 2$) is individually replaced by a
value randomly selected from a normal distribution with a mean
$\overline{\omega}_i = \omega_i$ and a standard deviation
$\sigma_{\omega_i} = \omega_i /25$. The statistical analysis of the
corresponding distributions of the fidelity and final-time entropy is
reported in table~\ref{tab:robust}, and frequency histograms of these
distributions are shown in figure~\ref{fig:RFO-f}. It is well known
\cite{RVM00, WaRa03, WRD93} that a high degree of quantum control may be
achieved through the complex interference of evolution pathways. This
interference strongly depends on the relative phases of all pathways,
and these phases in turn depend on the transition frequencies of the
system. Therefore, we would expect the optimal gate performance to be
much more sensitive to variations in the frequencies than to changes in
the coupling constants. The results presented in table~\ref{tab:robust}
and figure~\ref{fig:RFO-f} corroborate this expectation. Still, the
robustness of the optimal gate performance to frequency fluctuations is
tolerable. Moreover, the degree of robustness for systems with two and
more environmental particles ($n \geq 2$) is even higher than for $n =
1$.


\section{Conclusions}
\label{sec:conc}

This work demonstrates the importance of OCT in designing quantum gates
for use in QC, especially in the presence of a decohering
environment. The model studied here represents a realistic system of
interacting qubits and is relevant for various physical implementations
of QC. High quality optimal solutions obtained in the presence of
unwanted couplings also exhibit a significant degree of robustness to
random variations in the system parameters. The analysis of the system
dynamics reveals control mechanisms which employ fast and intense
time-dependent fields to effectively suppress the qubit-environment
interaction via dynamic shifting of the energy levels and achieve an
almost full coherence recovery via an induced revival.

The results reported in this paper further support the use in QC
applications of laboratory closed-loop optimal controls employing
learning algorithms and intense ultrafast fields \cite{RVM00,WaRa03}. In
the area of molecular dynamics, the utility of optimal control
methods was first demonstrated theoretically in very simple model
systems; nevertheless, these methods were later applied with great
success in the laboratory to complex molecules \cite{RVM00}. Similarly,
we expect that the optimal control of quantum gates, the usefulness of
which was demonstrated here for a relatively simple environment model,
will be also effective for real quantum information systems. A
successful application of optimal control methods to the generation
of high-fidelity quantum gates in the laboratory will be an important
step towards achieving error thresholds required for fault tolerant QC
\cite{FTQC,Threshold}.

This work may be further advanced with the use of the control-mechanism
analysis \cite{MR03} to explore the detailed dynamics of the decoherence
management process in optimally controlled quantum gates. Methods of
landscape analysis \cite{RHmR05, BRHWK06, RHR05} may be employed to
investigate how optimal controls are deduced and study the effects
of control errors in the context of non-unitary dynamics of open quantum
systems.

\ack

This work was supported by the ARO-QA, DOE, and NSF. D~A~L was supported
by ARO-QA Grant No.~W911NF-05-1-0440 and NSF Grant
No.~CCF-0523675. I~A~W acknowledges support by the UK QIP IRC funded by
EPSRC, and the EC under the Integrated Project QAP funded by the IST
directorate as Contract No.~015848.


\newpage


\end{document}